\documentclass[11pt,a4paper]{article}
\usepackage[margin=23mm,top=22mm,bottom=23mm]{geometry}
\usepackage[T1]{fontenc}
\usepackage[utf8]{inputenc}
\usepackage{newtxtext,newtxmath}
\usepackage{amsmath,bm}
\usepackage{graphicx}
\usepackage{microtype}
\usepackage{caption}
\usepackage[section]{placeins}
\usepackage{titlesec}
\usepackage[hidelinks,bookmarksnumbered]{hyperref}
\hypersetup{pdftitle={Nonreciprocal Floquet tunneling dynamics of spin-orbit-coupled bosonic atoms in a non-Hermitian double-well potential},pdfauthor={JU Limei et al.}}
\newcommand{\ket}[1]{\lvert #1\rangle}
\newcommand{\ii}{\mathrm{i}}
\newcommand{\dd}{\mathrm{d}}
\newcommand{\Bessel}{\mathcal{J}}
\DeclareMathOperator{\diag}{diag}
\titleformat{\section}{\large\bfseries}{\thesection.}{.6em}{}
\titleformat{\subsection}{\normalsize\bfseries}{\thesubsection.}{.6em}{}
\titleformat{\subsubsection}{\normalsize\bfseries}{\thesubsubsection.}{.6em}{}
\titlespacing*{\section}{0pt}{16pt}{7pt}
\titlespacing*{\subsection}{0pt}{12pt}{5pt}
\titlespacing*{\subsubsection}{0pt}{10pt}{5pt}
\begin{document}
\begin{center}
{\Large\bfseries Nonreciprocal Floquet tunneling dynamics of spin-orbit-coupled bosonic atoms in a non-Hermitian double-well potential\par}
\vspace{11pt}
JU Limei\textsuperscript{1)}\quad ZHANG Yingying\textsuperscript{1)\,$\dagger$}\quad
LIANG Jilei\textsuperscript{1)}\quad CHANG Shoukang\textsuperscript{1)}\\[4pt]
ZHU Zunlue\textsuperscript{1)}\quad QIN Lu\textsuperscript{1)\,$\ddagger$}\quad
LIU Wuming\textsuperscript{1)\,2)}\quad ZHAO Xingdong\textsuperscript{1)\,$\dagger\dagger$}
\vspace{8pt}

{\small\itshape
1) School of Physics, Henan Normal University, Xinxiang 453007, China\\[3pt]
2) Beijing National Laboratory for Condensed Matter Physics, Institute of Physics,\\
Chinese Academy of Sciences, Beijing 100190, China\par}
\vspace{6pt}
{\small (Received 8 February 2026; revised manuscript received 31 March 2026)}
\end{center}
\vspace{4pt}
\noindent\textbf{Abstract}\par
The stable tunneling dynamics in non-Hermitian quantum systems is one of the fundamental issues in the study of quantum transport. Nonreciprocal coupling, as another crucial manifestation of non-Hermiticity, exerts significant regulatory effects on the stability of quantum devices and on spin tunneling dynamics. Here we place spin-orbit-coupled bosonic atoms in a double-well potential with gain and loss, and investigate the effect of nonreciprocal coupling on the system dynamics. Through an analytical analysis of the Floquet states and the quasi-energy spectrum under periodic driving, we obtain the PT-symmetry phase diagram of the system. We further explore the synergistic regulation mechanism of several factors---including the strength of nonreciprocal coupling, the non-Hermitian strength arising from gain-loss, and the periodic driving---on stable spin-flip tunneling. The results show that precise manipulation of nonreciprocity and gain-loss can substantially enlarge the stable parameter region. In nonreciprocal systems with balanced gain and loss, the continuity or discreteness of the stable parameter region is determined by the parity of the driving-frequency ratio $\Omega/\omega$, and the spin-flip tunneling can be precisely controlled through the parameters $\lambda$ and $2\varepsilon/\omega$. For the unbalanced gain-loss scenario, the equilibrium conditions for achieving stable spin-flip tunneling under different parities of $\Omega/\omega$ are further provided. These findings offer a new theoretical approach for realizing robust spin-flip tunneling, provide guiding significance for the study of quantum-state transport in non-Hermitian ultracold atomic systems, and also serve as a reference for the experimental design of novel spin-based quantum devices.


\noindent\textbf{DOI:} 10.7498/aps.75.20260222

\section{Introduction}
In recent years, research on non-Hermitian physics in open quantum systems has come to cover a broad range of fields, including atomic and molecular physics, mesoscopic physics, and spin and magnetic systems [1-5]. A variety of novel physical phenomena have been explored together with it, including anomalous topological properties [6], localization transitions [7-9], the non-Hermitian skin effect [10-14], and spontaneous parity-time (PT) symmetry breaking [15], all of which have attracted wide research interest. In non-Hermitian systems, the complex energy spectrum associated with the Hamiltonian generally gives rise to unstable dynamics, which creates difficulties for the experimental observation of non-Hermitian phenomena [16]. As a bridge between the theory and experiment of non-Hermitian systems, PT-symmetric systems provide a real energy spectrum that guarantees the stability of the system [17]. 

At present, studies of non-Hermitian physics from various fields mostly consider the dissipation of open systems, or the presence of gain and loss in the system [18]. For example, in optical experimental systems, gain is introduced through a laser or other energy source, whereas loss arises from the attenuation of energy or a signal during propagation [19-21]; in a theoretical analysis one usually introduces the coupling between the system and the environment, or open boundary conditions, to realize gain and loss. Meanwhile, as an important branch of non-Hermitian physics, nonreciprocity has been studied in cavity-optomechanical systems, where optomechanically induced transparency [22-26], ground-state cooling of the mechanical resonator [27-29], and nonlinear effects [30-32], together with a series of novel physical phenomena, have been investigated [33]. In optical systems, nonreciprocity is usually implemented by introducing an artificial gauge field to construct asymmetric coupling [34,35], which is then used to explore the asymmetric response of forward and backward transmission or transition processes. This feature has key applications in optical circulators, topologically protected boundary states, and chiral quantum information processing [36-39]. Recently, loss-induced nonreciprocity has been proposed as a simple and convenient implementation and has been experimentally verified in artificial structures such as phononic crystals and circuits [40,41]. This scheme exploits the inherent energy dissipation of the system to break reciprocity, without requiring any external magnetic field or spatiotemporal modulation, and it provides a new idea for the design of nonreciprocal devices and for studying the topological properties of dissipative systems.

Recently, owing to advantages such as easy controllability, cold atoms have become one of the platforms for studying non-Hermitian physics. In 2019, the group of Prof. Luo Le at Sun Yat-sen University [42] observed PT symmetry breaking in a dissipative Floquet system of ultracold atoms. In 2022, the group of Prof. Yan Bo at Zhejiang University [43] realized an effective non-Hermitian SSH model using the momentum-lattice technique. In 2025, a research team at the Hong Kong University of Science and Technology [44] realized a quantum simulation of the two-dimensional non-Hermitian skin effect with ultracold fermions. In particular, research on spin-orbit-coupled quantum gases has achieved breakthrough progress in both theoretical construction and experimental observation [45-48], which provides new degrees of freedom for quantum-physics research based on ultracold atoms [49] and opens up new research directions. For example, the research team at Hunan Normal University [50-52], by periodically driving a non-Hermitian spin-orbit-coupled cold-atom double-well scheme, realized the manipulation of stable spin-flip tunneling; the extremely high controllable degrees of freedom of this system hold great potential for applications in quantum-information processing.

In this paper, we consider nonreciprocal tunneling between the wells of a non-Hermitian spin-orbit-coupled cold-atom double-well system, and study the effects of two non-Hermiticity-inducing mechanisms---gain-loss and nonreciprocity---on the system dynamics. Section 2 constructs the Hamiltonian of the system and solves it analytically on the basis of conventional Floquet theory. Section 3 analyzes, respectively, the phase diagrams and the transport dynamics of quantum states for nonreciprocal systems with balanced gain-loss and with unbalanced gain-loss. Finally, Section 4 gives a summary of the whole paper.

\section{Model and theory}
We study the nonreciprocal tunneling of a dipole ultracold boson trapped in a double well that carries gain and loss, as shown in Fig.~1(a), a pair of Raman lasers realizes effective spin-orbit coupling by coupling two internal atomic levels [53,54]. Here the two hyperfine internal states $\ket{1,-1}=\ket{\downarrow}$ and $\ket{1,0}=\ket{\uparrow}$ of $F=1$ of $^{87}\mathrm{Rb}$ atoms are chosen as the pseudospin states of the atoms, and they are kept well away from $\ket{1,+1}$ [55,56]. The dynamical Hamiltonian of this non-Hermitian system [57] is $H(t)=H_{\mathrm{Nrep}}(t)+H_{\mathrm{Zm}}(t)+H_{\mathrm{GL}}(t)$, where
\begin{align}
H_{\mathrm{Nrep}}(t)&=-\nu_1 a_l^{\dagger}\mathrm{e}^{-\ii\pi\lambda\hat{\sigma}_y}a_r
                     -\nu_2 a_r^{\dagger}\mathrm{e}^{\ii\pi\lambda\hat{\sigma}_y}a_l,\tag{1a}\\
H_{\mathrm{Zm}}(t)&=\frac{\Omega}{2}\sum_j\bigl(n_{j\uparrow}-n_{j\downarrow}\bigr),\tag{1b}\\
H_{\mathrm{GL}}(t)&=\sum_s\bigl[\varepsilon_l(t)n_{ls}-\varepsilon_r(t)n_{rs}\bigr].\tag{1c}
\end{align}
Here, in the Hamiltonian $H_{\mathrm{Nrep}}$, $\nu_{1,2}$ characterize the leftward (rightward) tunneling in the absence of spin-orbit coupling. In fact, one only needs to add a perturbing optical lattice to construct a superlattice system; the nonreciprocal hopping in the system is realized through an artificial gauge field, constituting the conventional Aubry-Andr\'e-Harper (AAH) model [57-59]. In this case one can define $\nu_1=\nu$ and $\nu_2=\nu(1-g)$, where $g$ denotes the tuning parameter of the left-right-well tunneling strength. When $g=0$ the transition between the two wells is reciprocal; to study the effect of nonreciprocity on the system dynamics, we focus on the case $g\ne0$. The vector operators $\hat{a}_j^{\dagger}=(\hat{a}_{j\uparrow}^{\dagger},\hat{a}_{j\downarrow}^{\dagger})$ and $\hat{a}_{js}^{\dagger}$ ($\hat{a}_{js}$) are the creation (annihilation) operators of an atom with pseudospin $s$ ($s=\uparrow,\downarrow$) in the $j$-th well ($j=l,r$ denotes the left and right wells). $\lambda$ denotes the effective spin-orbit-coupling strength, and $\hat{\sigma}_y$ is the $y$ component of the Pauli operator. In the Zeeman energy $H_{\mathrm{Zm}}$, $\Omega$ characterizes the effective Zeeman-field strength. The number operator $\hat{n}_{js}=\hat{a}_{js}^{\dagger}\hat{a}_{js}$ in the gain-loss term $H_{\mathrm{GL}}$ counts the number of particles with spin $s$ in the $j$-th well. Gain and loss are realized by the periodic drive $\varepsilon_j(t)=\varepsilon\cos(\omega t)+\ii\beta_j$, with $\beta_l=\beta$ and $\beta_r=\alpha\beta$. Experimentally, gain and loss can be achieved by applying a focused electron beam between the condensates [60,61].

\begin{figure}[!htbp]
\centering
\includegraphics[width=0.6\linewidth]{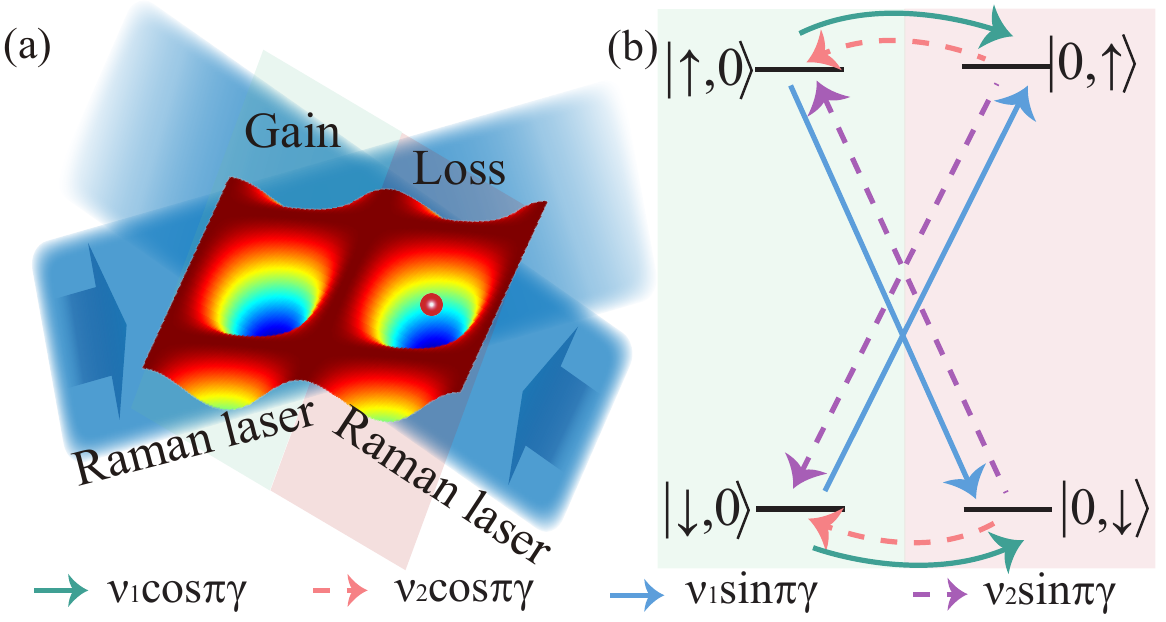}
\caption{(a) Illustrates a non-Hermitian double-well model constructed by a pair of counterpropagating Raman beams, containing a spin-orbit-coupled ultracold boson within the wells. Here, the green and red regions correspond to the gain and loss of the left and right wells, respectively; (b) Depicts the transition processes between different states of the system, where the green solid and orange dashed lines represent interwell tunneling without spin flip, while the blue solid and purple dashed lines represent interwell tunneling with spin flip. $\nu_1$ and $\nu_2$ denote the tunneling strengths between the two wells in the absence of spin-orbit coupling. The black horizontal solid lines indicate the four modes corresponding to their respective positions.}
\label{fig:1}
\end{figure}

Considering single-particle occupation of the system, i.e. using the Fock basis $\ket{0,s}$ ($\ket{s,0}$) to denote the state in which an atom with spin $s$ resides, expanding the wavefunction in this basis gives
\begin{equation}
\ket{\psi_a(t)}=a_1(t)\ket{0,\uparrow}+a_2(t)\ket{\downarrow,0}
              +a_3(t)\ket{\uparrow,0}+a_4(t)\ket{0,\downarrow},\tag{2}
\end{equation}
where $a_k(t)$ ($k=1,2,3,4$) is the probability amplitude of the atom in the corresponding state, with probabilities $P_k(t)=\lvert a_k(t)\rvert^2$. Fig.~1(b) is a schematic of the particle tunneling among the four quantum states. To obtain an analytical solution we impose the restriction $\omega\gg\nu,\beta_j$; $\Omega=n\omega$ ($n$ is an integer) [62], and then introduce the transformation
\begin{equation}
\bigl[a_1(t),a_2(t),a_3(t),a_4(t)\bigr]=\bigl[b_1(t),b_2(t),b_3(t),b_4(t)\bigr]U.\tag{3}
\end{equation}
Here the transformation matrix is
\[
\begin{aligned}
U=\diag\Bigl\{&
\exp\!\left[-\ii\int\!\left(\frac{\Omega}{2}-\varepsilon\cos\omega t\right)\dd t\right],\;
\exp\!\left[-\ii\int\!\left(-\frac{\Omega}{2}+\varepsilon\cos\omega t\right)\dd t\right],\\
&\exp\!\left[-\ii\int\!\left(\frac{\Omega}{2}+\varepsilon\cos\omega t\right)\dd t\right],\;
\exp\!\left[-\ii\int\!\left(-\frac{\Omega}{2}-\varepsilon\cos\omega t\right)\dd t\right]\Bigr\}.
\end{aligned}
\]
Finally, the transformation coefficients can be written through a Fourier expansion as
\begin{align*}
\exp\!\left[\pm\ii\int 2\varepsilon\cos\omega t\,\dd t\right]
 &=\sum_{n'=-\infty}^{\infty}J_{n'}\!\left(\frac{2\varepsilon}{\omega}\right)
   \exp(\pm\ii n'\omega t),\\
\exp\!\left\{\pm\ii\int\bigl[2\varepsilon\cos\omega t\pm\Omega\bigr]\dd t\right\}
 &=\sum_{n''=-\infty}^{\infty}J_{n''}\!\left(\frac{2\varepsilon}{\omega}\right)
   \exp\!\left\{\pm\ii\left(n''\pm\frac{\Omega}{\omega}\right)\omega t\right\}.
\end{align*}
Neglecting the fast-oscillating terms yields the effective undriven model
\begin{equation}
\ii\frac{\dd}{\dd t}
\begin{pmatrix}b_1(t)\\b_2(t)\\b_3(t)\\b_4(t)\end{pmatrix}
=\begin{pmatrix}
-\ii\beta_r&-J_4&-J_2&0\\
-J_3&\ii\beta_l&0&-J_1\\
-J_1&0&\ii\beta_l&J_5\\
0&-J_2&J_6&-\ii\beta_r
\end{pmatrix}\begin{pmatrix}b_1(t)\\b_2(t)\\b_3(t)\\b_4(t)\end{pmatrix}.\tag{4}
\end{equation}
Here the effective coupling constants are
\[
\begin{aligned}
J_{1,2}&=\nu_{1,2}\cos(\pi\lambda)\,\Bessel_0\!\left(\frac{2\varepsilon}{\omega}\right),\\
J_{3,4}&=\nu_{1,2}\sin(\pi\lambda)\,\Bessel_{\Omega/\omega}\!\left(\frac{2\varepsilon}{\omega}\right),\\
J_{5,6}&=\nu_{1,2}\sin(\pi\lambda)\,\Bessel_{-\Omega/\omega}\!\left(\frac{2\varepsilon}{\omega}\right),
\end{aligned}
\]
where $\Bessel_n(x)$ is the $n$-th order Bessel function. When $n$ is even, $\Bessel_n(x)=\Bessel_{-n}(x)$; when $n$ is odd, $\Bessel_n(x)=-\Bessel_{-n}(x)$. Clearly, the effective coupling constants can be controlled by tuning the system parameters.

According to conventional Floquet theory, we first find the stationary solution of Eq.~(4), $\ket{\psi_b(t)}=(A,B,C,D)^{T}\mathrm{e}^{-\ii Et}$. Then, according to the transformation (3), we obtain the solution of the system $\ket{\psi_a(t)}=\ket{\varphi_a(t)}\mathrm{e}^{-\ii Et}$, where $\ket{\varphi_a(t)}=(A,B,C,D)U$ is the Floquet state of the system, with the corresponding eigenvalue $E$, whose analytical expression is
\begin{equation}
E_{1,2}=\frac{\ii}{2}\left(\beta_l-\beta_r\pm\rho_-\right),\qquad
E_{3,4}=\frac{\ii}{2}\left(\beta_l-\beta_r\pm\rho_+\right),\tag{5}
\end{equation}
where 
\[
\rho_\pm=\sqrt{2\bigl(-2J_1J_2-J_3J_4-J_5J_6\pm\ii Q\bigr)+(\beta_l+\beta_r)^2},\] 
with 
\[
Q=\sqrt{\left [ -J_3^2J_4^2+4J_2^2J_3J_5+4J_1^2J_4J_6+2J_3J_4J_5J_6-J_5^2J_6^2-4J_1J_2\left ( J_3J_4+J_5J_6 \right )  \right ] }.
\]
From the above eigenvalues one obtains the four Floquet states $\ket{\varphi_a(t)}$, and the fundamental solutions $\ket{\psi_a(t)}$ of the periodically driven system can be further found.

\section{Results and discussion}
Based on solving the above coupled equations and on a stability analysis, the quasi-energies of the Floquet states can be written as the sum of a real part and an imaginary part, $E_p=\operatorname{Re}(E_p)+\mathrm{i}\operatorname{Im}(E_p)$ ($p=1,2,3,4$), as in Eq. (5), and the stability depends on the imaginary part of the quasi-energy, $\operatorname{Im}(E_p)$. When all $\operatorname{Im}(E_p)=0$, the quasi-energies are purely real, the evolution probabilities of the particle states vary periodically with time, and the system is stable. Searching for the region in which the quasi-energy is real corresponds to finding the stable phase region; this can be regarded as a special case of non-Hermitian systems, namely PT-symmetric systems. Combining the PT-symmetry criterion $\mathcal{PT}H=H\mathcal{PT}$, one can see that the system satisfies PT symmetry, so finding the stable parameter region is equivalent to computing the PT-symmetric phase region. Meanwhile, the threshold between PT symmetry and PT symmetry breaking is the transition between the stable and unstable regions, i.e. the position at which a non-Hermitian degeneracy point, or exceptional point (EP), appears. In addition, if part of $\operatorname{Im}(E_p)$ equals zero while another part is less than zero, then because the total probability of the particle-state evolution approaches a constant at very long times, the system can also be regarded as stable. Otherwise, the particle-state evolution probability grows or decays exponentially, i.e. the system is unstable.

Given that the system possesses gain and loss, we study the dynamics of the non-Hermitian system for two cases: balanced gain-loss (i.e. the gain-loss coefficients of the two wells are equal, $\alpha=1$) and unbalanced gain-loss (i.e. the gain-loss coefficients of the two wells are unequal, $\alpha\ne 1$). In addition, spin-flip tunneling is necessarily accompanied by an energy change $\Omega(-\Omega)$; when the ratio of the Zeeman-field strength to the driving frequency is an integer ($\Omega/\omega$), the system can exchange energy with the driving field to compensate for the energy loss caused by spin-flip tunneling. By Floquet averaging theory, the periodically driven spin-orbit-coupled system can be transformed into an effectively equivalent and more tractable undriven (static) system. In this case, the tunneling probability between the states $\ket{\downarrow ,0}$ and $\ket{0,\uparrow}$ can be characterized by $\mathcal{J}_{n}(2\varepsilon/\omega)$, while the tunneling probability between the states $\ket{\uparrow ,0}$ and $\ket{0,\downarrow}$ can be characterized by $\mathcal{J}_{-n}(2\varepsilon/\omega)$. Whether $\mathcal{J}_{n}(2\varepsilon/\omega)$ and $\mathcal{J}_{-n}(2\varepsilon/\omega)$ are equal depends on the parity of $n=\Omega/\omega$, from which one can further discuss whether the flip tunneling probabilities between the corresponding states are equal. Therefore, in this system one can study not only the effects of the non-Hermitian and nonreciprocal strengths on the inter-state tunneling probability, but also, by combining the type of $n$, analyze the influence of the corresponding quasi-energy spectrum on the stable region under different parameters.

\begin{figure}[!htbp]
\centering
\includegraphics[width=0.7\linewidth]{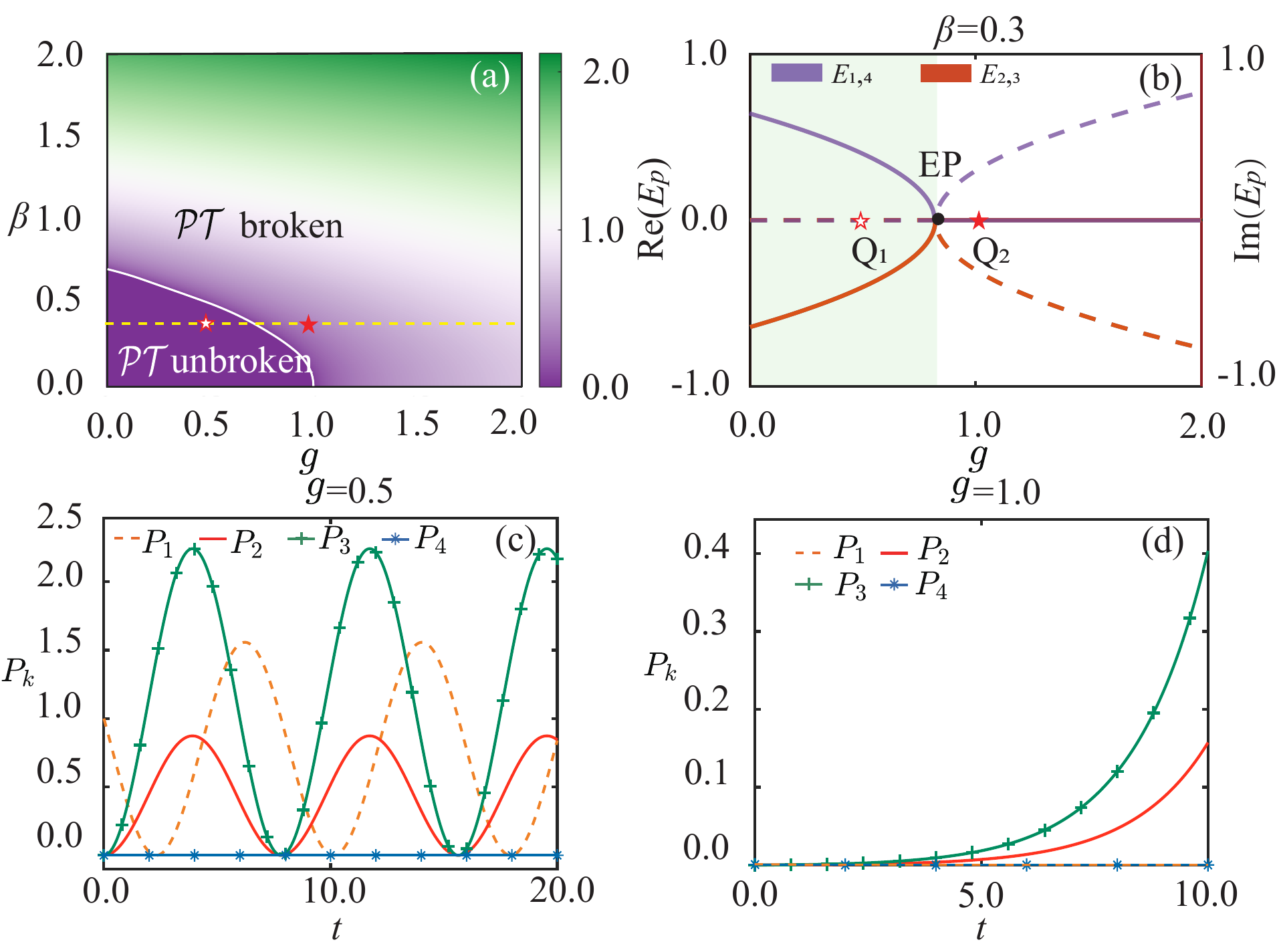}
\caption{(a) Quasi-energy $\operatorname{Im}(E_p)$ as a function of the balanced non-Hermitian strength $\beta$ and the nonreciprocal tuning parameter $g$; the white curve is the boundary between the regions where $\operatorname{Im}(E_p)$ is zero and nonzero. (b) At $\beta=0.3$, the real part $\operatorname{Re}(E_p)$ (solid) and imaginary part $\operatorname{Im}(E_p)$ (dashed) of the quasi-energy versus $g$; EP denotes the phase-transition point between PT symmetry and PT symmetry breaking. (c),(d) For an initial state $\ket{0,\uparrow}$, the time evolution of the probabilities of different states for $g$ in the stable region ($Q_{1}$) and the unstable region ($Q_{2}$). Other parameters: $\lambda=0.3, x=2\varepsilon/\omega=1.5, \alpha=1$.}
\label{fig:2}
\end{figure}

\subsection{Nonreciprocal system with balanced gain and loss}
\subsubsection{The case in which \texorpdfstring{$\Omega/\omega$}{Omega/omega} is even}

When $\alpha=1$, i.e. $\beta _{r}=\beta _{l}=\beta$, combining Eq. (5), the imaginary part of the quasi-energy $E_{p}$ depends on whether $\operatorname{Re}(\rho_\pm)$ exists, and the form of $\rho _{\pm}$ is related to the effective coupling constants $J_{1\text{-}6}$. We first discuss how the quasi-energy and the distribution of the stable region vary with the parameters when $\Omega/\omega$ is even and when it is odd. When $\Omega/\omega=2$ ($\Omega=100, \omega=50$) is even, Fig. 2(a) shows the distribution of the quasi-energy $\operatorname{Im}(E_p)$ as a function of $g$ and $\beta$. In the figure, the purple region represents the stable region corresponding to $\operatorname{Im}(E_p)=0$, and the white curve is the boundary at which the quasi-energy acquires an imaginary part. Clearly, when $g=0$ (the reciprocal case) the system does not begin to transition into the unstable region until $\beta$ increases to 0.713; further increasing $g$ makes the system nonreciprocal, and the critical value of $\beta$ for the unstable region decreases. Similarly, increasing $\beta$ gradually decreases the critical value of $g$ for the unstable region. Therefore, by setting $\beta(g)$ to a reasonable value one can achieve a larger tunable stable $g(\beta)$ parameter interval. Meanwhile, to verify the energy distribution on both sides of the white curve, we select $\beta=0.3$ to draw Fig. 2(b). One can see that when $g=0.817$ the quasi-energy changes from purely real to purely imaginary, i.e. the EP. The white curve in Fig. 2(a) is the EP line formed by connecting several EPs; the two sides of the EP line correspond respectively to the PT-symmetric and the PT-symmetry-broken phase regions, i.e. the stable and unstable regions of the system.

To further explore the spin-flip tunneling in the different phase regions, taking points $Q_{1}$ and $Q_{2}$ in Fig. 2(b) as examples, we plot the evolution with time of the probability of each state, $P_k=\lvert a_k(t)\rvert^2$, as shown in Fig. 2(c) and 2(d). One can see that in Fig. 2(c), when the initial state of the particle is $\ket{0,\uparrow}$, after evolving for a time $t$ it transitions to $\ket{\downarrow ,0}$ with a certain probability; here stable spin-flip tunneling occurs, but it is accompanied by an oscillating and non-conserved total probability---a generalized Rabi oscillation. Similarly, the particle also transitions to $\ket{\uparrow ,0}$ with a certain probability, which corresponds to a stable Rabi oscillation without spin flip. Under the same initial conditions, Fig. 2(d) gives the evolution of the system in the PT-symmetry-broken phase. One can see that unstable spin tunneling can cause the state-evolution probability to grow without bound, exponentially. This result shows that increasing the nonreciprocal tuning parameter $g$ turns stable spin-flip tunneling into unstable spin tunneling, and that increasing $g$ preferentially suppresses stable spin-flip tunneling.

In non-Hermitian spin-orbit-coupled cold-atom systems, how to effectively manipulate spin transport within the stable region is a much-studied issue, and so here we tune the system parameters in multiple dimensions. We study the effect of continuously varying the spin-orbit-coupling strength $\lambda$ and $2\varepsilon/\omega$ on the stable quasi-energy region. When $g=0.5, \nu=2, \beta=0.3$, the system repeatedly corresponds to stable states over continuous parameter regions, as shown in Fig. 3(a). The widths of these continuous stable parameter regions vary periodically with $\lambda$ and narrow as $2\varepsilon/\omega$ increases. Further increasing $\beta$ to 0.6, the continuous stable regions gradually shrink until they disappear, and discrete stable parameter regions appear, as shown in Fig. 3(b). To further explore how the values of $\lambda$ and $2\varepsilon/\omega$ in the stable region affect the particle's spin-flip tunneling, we select stable points ($\mathrm{Q_{1}}, \mathrm{Q_{2}}, \mathrm{Q_{3}}$) from Figs. 3(a)-(b), and show the time-evolved state probabilities in Figs. 4 and 5.

\begin{figure}[!htbp]
\centering
\includegraphics[width=0.68\linewidth]{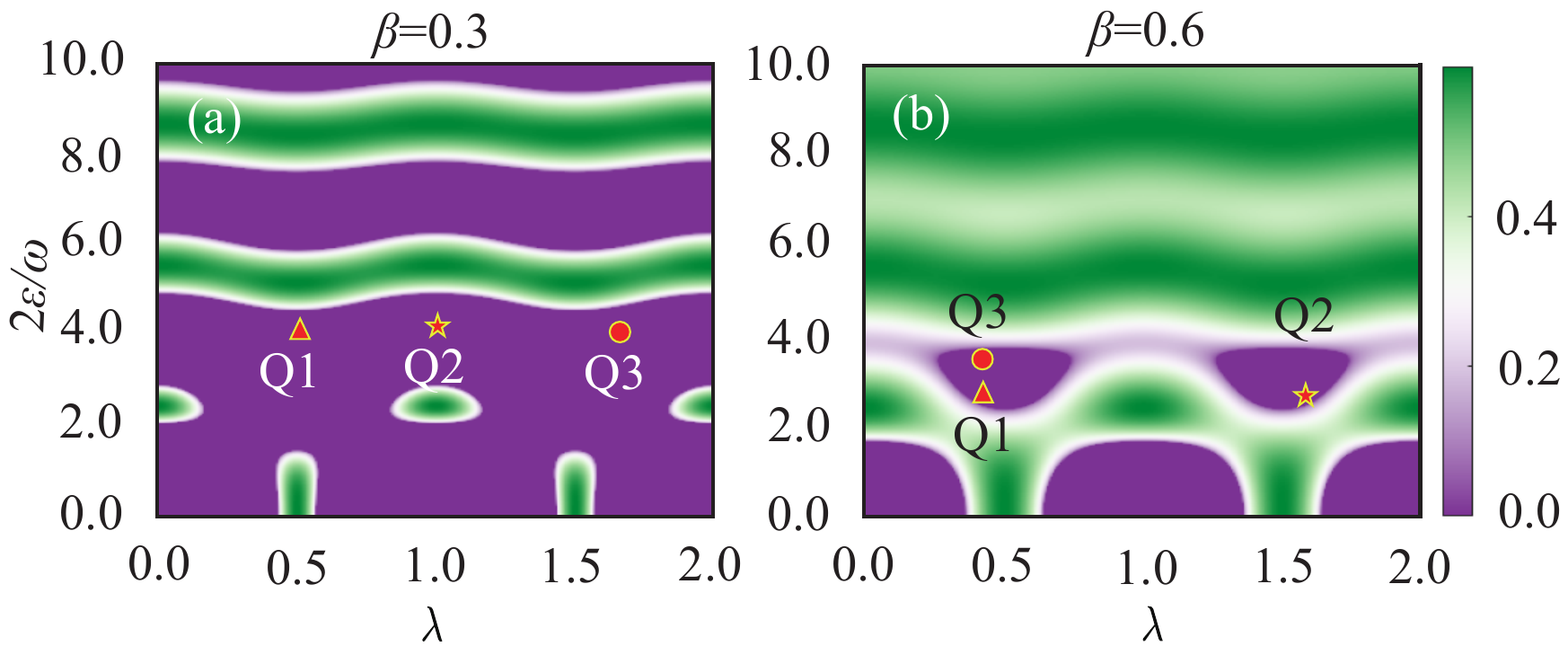}
\caption{(a) and (b) show the distribution of $\operatorname{Im}(E_p)$ as $\lambda$ and $2\varepsilon/\omega$ vary continuously, for $\beta=0.3$ and 0.6, respectively. The boundary between the purple and green regions marks the parameter threshold separating the stable and unstable zones. Points $\mathrm{Q_{1}}$, $\mathrm{Q_{2}}$, and $\mathrm{Q_{3}}$ are selected in the stable regions to study the spin-flip tunneling behavior of the particle. The other parameters are dimensionless: $\nu=2, g=0.5, \Omega=100, \omega=50$.}
\label{fig:3}
\end{figure}

Figures 4(a)-(c) show the variation of $\operatorname{Re}(E_p)$ and $\operatorname{Im}(E_p)$ as $2\varepsilon/\omega$ increases for $\lambda=0.5$, $1.0$, and $1.7$, respectively. The critical points at which the real and imaginary parts alternate are the phase-transition points from PT symmetry to breaking, with the corresponding phase regions colored green and white. To observe how different $\lambda$ values change the time evolution of the system-state distribution within the same stable region, we select points $\mathrm{Q_{1}}, \mathrm{Q_{2}}, \mathrm{Q_{3}}$ in the PT-symmetric phase region, as shown in Figs. 4(d)-(f). When $\lambda=0.5$ (Fig. 4(d)), the system coupling constants satisfy $J_{1}=J_{2}=0$, leaving only the spin-flip-related coupling channels open, and only spin-flip quantum tunneling occurs. When $\lambda$ increases to 1 (Fig. 4(e)), the coupling constants show another feature: $J_{3}=J_{4}=0$, i.e. all spin-flip-related coupling channels are closed, and only non-spin-flip quantum tunneling occurs. When the value of $\lambda$ makes neither $J_{1}=J_{2}=0$ nor $J_{3}=J_{4}=0$, stable generalized spin-flip and non-spin-flip Rabi oscillations occur simultaneously, as shown in Fig. 4(f). Thus, by adjusting the spin-orbit-coupling strength $\lambda$ one can precisely control the particle's tunneling process.

\begin{figure}[!htbp]
\centering
\includegraphics[width=0.85\linewidth]{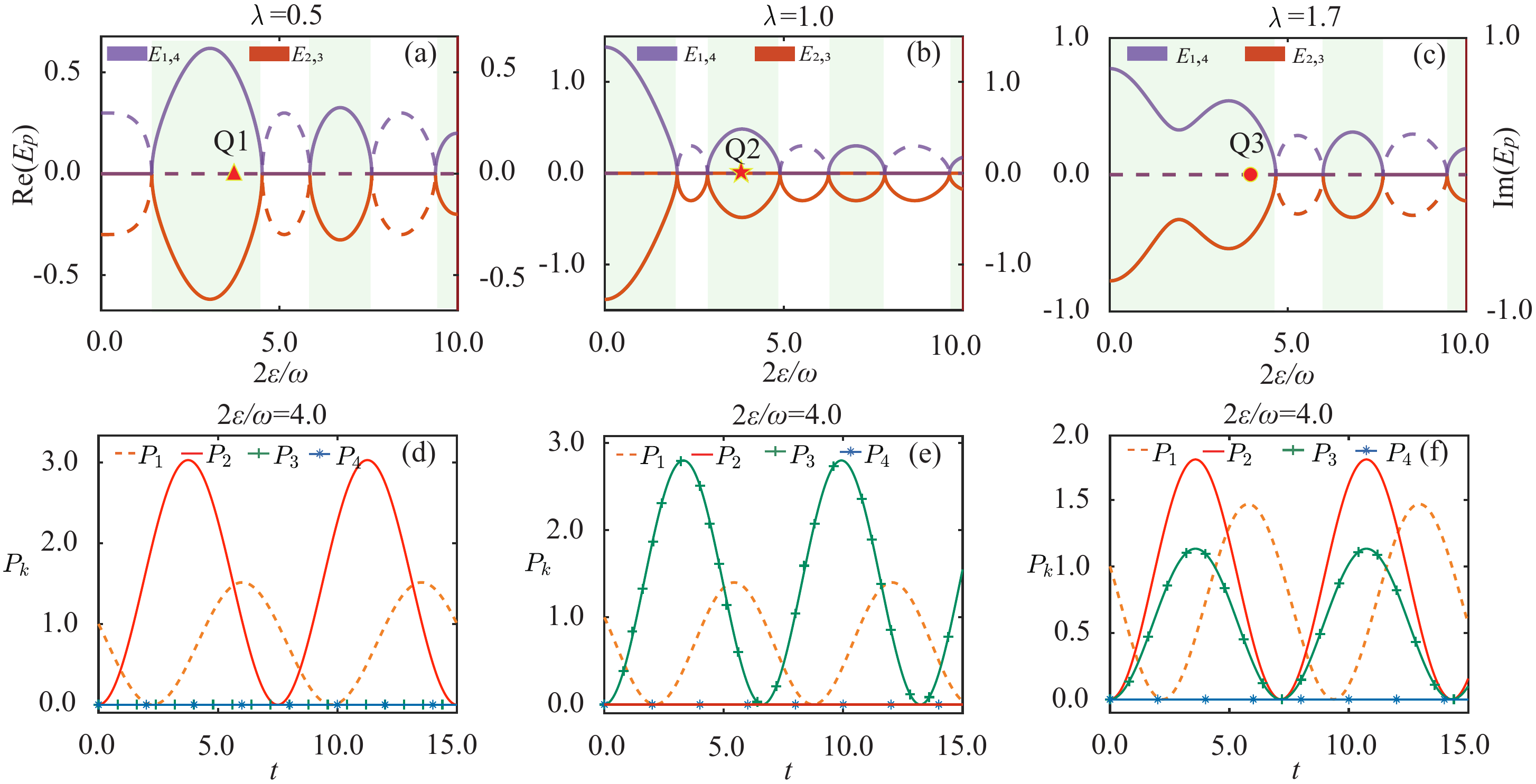}
\caption{The first row shows the real part $\operatorname{Re}(E_p)$ (solid lines) and imaginary part $\operatorname{Im}(E_p)$ (dashed lines) of the system quasi-energy as functions of $2\varepsilon/\omega$ for spin-orbit-coupling strengths (a) $\lambda=0.5$, (b) $\lambda=1.0$, and (c) $\lambda=1.7$, respectively. The second row shows, under the condition $2\varepsilon/\omega=4$, the time evolution of the particle-state probabilities at $\mathrm{Q_{1}}$ (d), $\mathrm{Q_{2}}$ (e), and $\mathrm{Q_{3}}$ (f) for the different stable parameter regions of the first row. Here the initial state of the particle is $\ket{0,\uparrow}$.}
\label{fig:4}
\end{figure}

Figures 5(a)-(c) show the variation of $\operatorname{Re}(E_p)$ and $\operatorname{Im}(E_p)$ as $\lambda$ increases for $2\varepsilon/\omega=2.405$ and 3.0, respectively. Similar to the analysis of Fig. 4, in each panel we select a stable point to study the corresponding evolution of the state probabilities with time. We find that adjusting the value of $2\varepsilon/\omega$ can directly control the occurrence of spin flip, and increasing $2\varepsilon/\omega$ can also put the system in a state in which spin-flip and non-spin-flip coexist. This is because when $2\varepsilon/\omega=2.405$ the zero-order Bessel function of the first kind satisfies the zero condition $\mathcal{J}_{0}(2\varepsilon/\omega)=0$ (where $\mathcal{J}_{0}(x)$ is the zero-order Bessel function of the first kind). Under this condition $J_{1}=J_{2}=0$ (the non-spin-flip-related coupling channels are closed), while $J_{3},J_{4}\ne 0$ (the spin-flip-related coupling channels remain open), so the system only allows spin-flip quantum tunneling. When $\lambda=0.4$ and $\lambda=1.5$, only the stable generalized Rabi oscillation involving spin flip occurs, with time-evolution curves shown in Figs. 5(d) and 5(e). For Fig. 5(c) ($2\varepsilon/\omega=3$), the Bessel function does not satisfy any zero condition, so all coupling channels remain open, i.e. $J_{1},J_{2}\ne 0$ and $J_{3},J_{4}\ne 0$. This feature directly leads to the simultaneous presence of non-flip spin tunneling (corresponding to the $J_{1},J_{2}$ channels) and flip spin tunneling (corresponding to the $J_{3},J_{4}$ channels); the competition and cooperation between the two tunneling processes make the population-probability evolution exhibit more complex oscillatory features, as shown in Fig. 5(f). Thus, adjusting $2\varepsilon/\omega$ changes the flip character of the particle tunneling and allows better control of the particle's tunneling.

\begin{figure}[!htbp]
\centering
\includegraphics[width=0.85\linewidth]{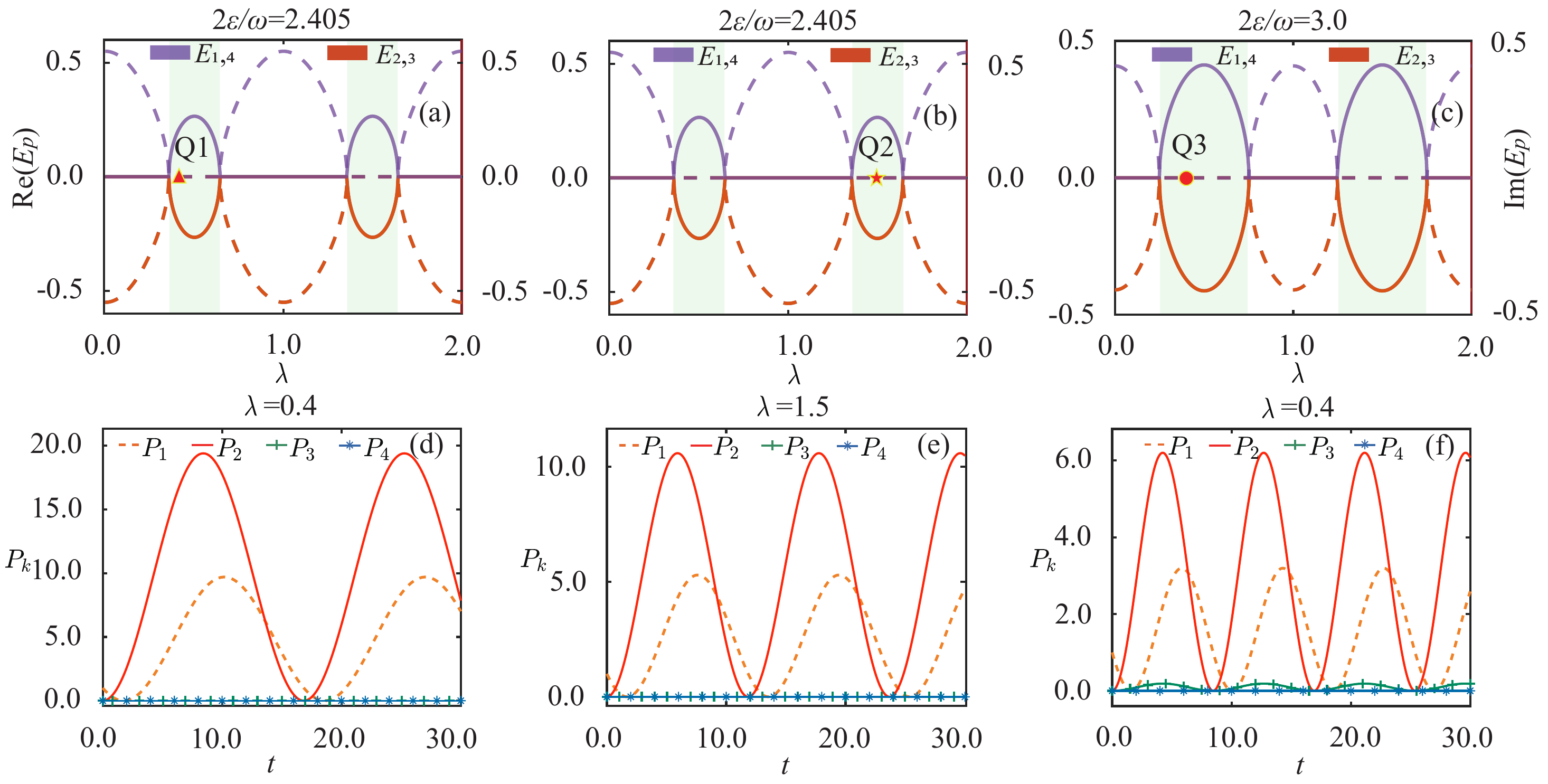}
\caption{The first row shows the real part $\operatorname{Re}(E_p)$ (solid lines) and imaginary part $\operatorname{Im}(E_p)$ (dashed lines) of the system quasi-energy as functions of $\lambda$, for $2\varepsilon/\omega$ equal to (a) 2.405, (b) 2.405, and (c) 3.0, respectively. The second row shows the time evolution of the particle-state probabilities at $\mathrm{Q_{1}}, \mathrm{Q_{2}}$, and $\mathrm{Q_{3}}$, corresponding to the different stable parameter regions of the first row; $\mathrm{Q_{1}}$ and $\mathrm{Q_{3}}$ correspond to $\lambda=0.4$ ((d),(f)) and $Q_{2}$ corresponds to $\lambda=1.5$ (e). The initial particle state is $\ket{0,\uparrow}$.}
\label{fig:5}
\end{figure}
In summary, when $\Omega/\omega$ is even the stable region and the PT-symmetric phase can be manipulated by properly tuning $\beta$ and $g$. On the other hand, in the tunable parameter space of $\lambda$ and $2\varepsilon/\omega$, based on the periodicity of the effective coupling constants $J_{1\text{-}6}$ and on the zeros of the Bessel functions, one finds that only when $\lambda$ equals an odd multiple of $1/2$ does $J_{1}=J_{2}=0$ (the non-spin-flip coupling channels closed); and only when $\lambda$ equals an even multiple of $1/2$ does $J_{3}=J_{4}=0$ (the spin-flip coupling channels closed). Meanwhile, as long as the Bessel function satisfies the zero condition (i.e. $2\varepsilon/\omega=2.405$), $J_{1}=J_{2}=0$ and only the spin-flip-related coupling channels are open; otherwise all coupling channels are open.

\subsubsection{The case in which \texorpdfstring{$\Omega/\omega$}{Omega/omega} is odd}
In the other case, when $\Omega/\omega=1$ ($\Omega=\omega=50$) is odd, the conditions $J_{3}=-J_{5}$ and $J_{4}=-J_{6}$ hold. In the balanced gain-loss nonreciprocal system, the parameter $\rho$ in the quasi-energy $E_{p}$ expression in Eq. (5) can be rewritten as \[\rho_\pm=2\sqrt{\beta^2-(J_1J_2+J_3J_4)\pm\lvert J_2J_3+J_1J_4\rvert}.\] We present the phase diagram based on the stability analysis in this case, as shown in Fig. 6(a), giving the PT-symmetry phase diagram in the $g$-$\beta$ plane. Compared with the even case, the stable phase region is obviously reduced. It is clear that the parity of $\Omega/\omega$ has a significant effect on the stability of the system. To study the spin-flip tunneling in the stable region of Fig. 6(b), we take $\beta=0.2$ and analyze the influence of $g$ on the quasi-energy. As shown in Fig. 6(b), the first EP at $g=0.56$ serves as the critical point controlling the system stability; before and after it, the state-probability evolution still corresponds to stability and instability. The difference is that here the initial state ($\ket{0,\uparrow}$) in the stable parameter region evolves with time into a richer set of particle states (Fig. 6(c)); this numerical result agrees with the analytical analysis of Eq. (4).

For a fixed nonreciprocal strength, we analyze the stability in the parameter space ($\lambda, 2\varepsilon/\omega$). We find that the stable regions in this parameter space all become discrete, as shown in Fig. 6(d) and 6(e). Obviously, the stability boundary curves correspond to integer multiples of $\lambda=1/2$. Further analysis shows that the parameter width of the discrete stable regions gradually decreases as $2\varepsilon/\omega$ increases (Fig. 6(f)) and varies periodically with $\lambda$ (Fig. 6(g)). This indicates that the specific tunneling channels of the system are blocked and quantum spin tunneling is not allowed. This is because, during the variation of $2\varepsilon/\omega$ or $\lambda$, $(J_1J_2+J_3J_4)\pm\lvert J_2J_3+J_1J_4\rvert$ may be zero, so that when $\beta$ is nonzero $\operatorname{Re}(\rho_-)=0$ cannot hold. Increasing $\beta$ (Fig. 6(e)) further compresses the stable regions. Next, taking points $\mathrm{Q_{1}}$ ($\lambda=1, 2\varepsilon/\omega=2$) and $\mathrm{Q_{2}}$ ($\lambda=0.4, 2\varepsilon/\omega=2.405$) in Fig. 6(f) and 6(g), we further analyze the evolution of the system with time for a specified initial state. Combined with the even case, the analysis of special points of ($\lambda, 2\varepsilon/\omega$) also shows that when $\lambda$ and $2\varepsilon/\omega$ take different special values in the same stable region, the spin-flip character of the particle state is completely different; this result is similar to the previous one (Figs. 6(h) and 6(i)). This also proves that the parity of $\Omega/\omega$ and the nonreciprocal strength only significantly affect the distribution of the stable region, whereas the spin-flip tunneling in the stable region is mainly determined by $\lambda$ and $2\varepsilon/\omega$.

\begin{figure}[!htbp]
\centering
\includegraphics[width=0.90\linewidth]{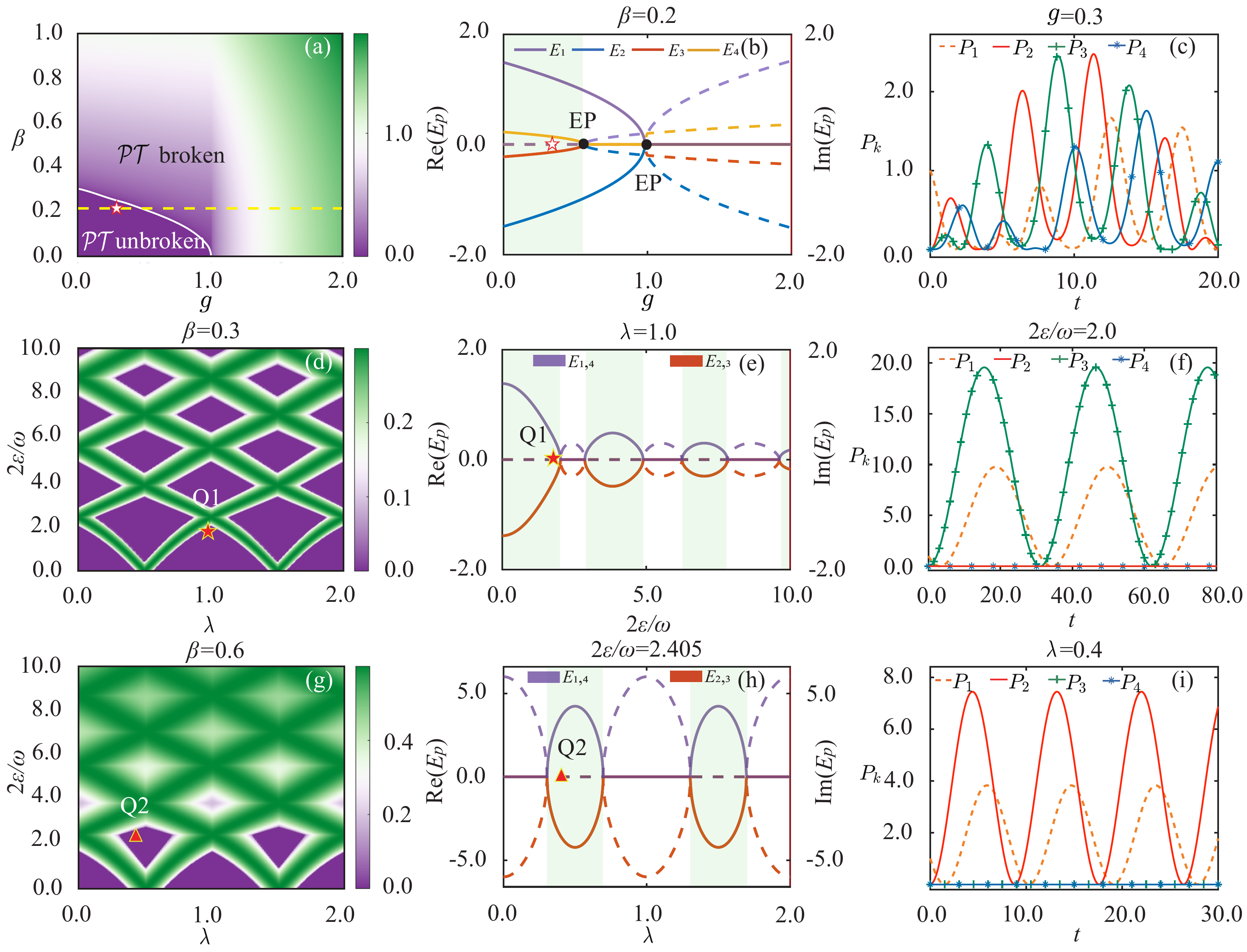}
\caption{For $\lambda=0.3, 2\varepsilon/\omega=1.5$, and odd $\Omega/\omega$. (a) Distribution of the imaginary part of the quasi-energy, $\operatorname{Im}(E_p)$, as a function of $\beta$ and $g$; the solid white line is the boundary between the PT-symmetric and PT-symmetry-breaking phase transitions. (b) Taking $\beta=0.2$, the real part (solid) and imaginary part (dashed) of the eigenenergy versus $g$; the eigenenergy degenerates twice, at $g=0.556$ and 1.0, forming two exceptional points (EPs). (c) In the stable region before the first EP, taking $g=0.3$, the time evolution of the probabilities of different states for the initial state $\ket{0,\uparrow}$. (d),(e) Distribution of $\operatorname{Im}(E_p)$ as $\lambda$ and $2\varepsilon/\omega$ vary continuously for $\beta=0.3$ and 0.6; the white boundaries separate the stable (purple) and unstable (green) regions. Points $\mathrm{Q_{1}}$ ($\lambda=1$) and $\mathrm{Q_{2}}$ ($2\varepsilon/\omega=2.405$) are selected to plot $E_{p}$ versus the other parameter in (f),(g). The time evolution at $\mathrm{Q_{1}},\mathrm{Q_{2}}$ for the initial state $\ket{0,\uparrow}$ is shown in (h),(i).}
\label{fig:6}
\end{figure}

\subsection{Nonreciprocal system with unbalanced gain and loss}

In fact, studying non-Hermitian physics by tuning gain and loss is currently a research hotspot in many fields. In our system, the unbalanced gain-loss case is realized simply by taking different gain-loss strengths for the two wells, i.e. $\beta _{r}\ne \beta _{l}$ (or $\alpha\ne 1$).

From the previous analysis, when $\Omega/\omega$ is even, $\rho _{\pm}$ in Eq. (5) can be expressed as $\sqrt{(\beta_l+\beta_r)^2-4(J_1J_2+J_3J_4)}$. Solving for the quasi-energy of the system, under the condition of stability one obtains the relative relation between the effective coupling strength and the gain-loss coefficients, i.e. $\beta _{l}\beta _{r}=J_{1}J_{2}+J_{3}J_{4}$, with $E_{1}=E_{4}=0$ and $E_2=E_3=\mathrm{i}(\beta_l-\beta_r)$. To further verify the analysis, we fix some parameters $\Omega=100, \omega=50, \lambda=0.3, \varepsilon=37.5, g=0.5, \nu=2$, and, using the relation between $\beta _{l}$ and $\beta _{r}$ under the stability condition, plot the time evolution of the particle-state probabilities. As shown in Fig. 7, here $P=\sum_{k=1}^{4}P_k$. Figures 7(a) and 7(b) show the time-evolved probability distribution of particles in different initial states under the same parameters. One can see that whether the particle is initially in the left or right well, the total probability changes rapidly within a short time interval and gradually reaches a stable value after some evolution. Namely, if the particle starts in the right (loss) well, the total probability decreases to a stable value of 0.49 (Fig. 7(a)); the initial state $\ket{0,\uparrow}$ tunnels to the left well with a relatively larger probability, corresponding to $\ket{\uparrow ,0}$. However, when the particle initially resides in the left (gain) well, its total probability increases to a higher stable value of about 1.51 (Fig. 7(b)), and its initial state $\ket{\uparrow ,0}$ undergoes spin flip with a larger probability.

In the other case, when $\Omega/\omega$ is odd, the quasi-energy changes. Combined with the above analysis, the four quasi-energies make the system stable under two conditions. First, one quasi-energy is zero and the imaginary parts of the others are all less than zero. Second, two quasi-energies are zero and the imaginary parts of the other two are less than zero. The first case corresponds to two relations:
\[
\begin{aligned}
\text{(1)}\quad &\beta_l\beta_r=(J_1J_2-J_3J_4)-\lvert J_2J_3+J_1J_4\rvert,\\
& (\beta_l+\beta_r)^2<4\bigl[(J_1J_2+J_3J_4)+\lvert J_2J_3+J_1J_4\rvert\bigr];\\[5pt]
\text{(2)}\quad &\beta_l\beta_r=(J_1J_2-J_3J_4)-\lvert J_2J_3+J_1J_4\rvert,\\
& 0\le\rho_+\le\beta_r-\beta_l.
\end{aligned}
\]
Taking (1) as an example, the quasi-energies are
\[
\begin{aligned}
E_1&=0, & E_2&=\ii(\beta_l-\beta_r),\\
E_3&=\frac{\ii}{2}(\beta_l-\beta_r-\rho_+), &
E_4&=\frac{\ii}{2}(\beta_l-\beta_r+\rho_+).
\end{aligned}
\]
When $\beta _{r}>\beta _{l}$, the purely imaginary character of $\rho _{+}$ brings the system to a stable state. Under the parameter conditions (Fig. 7(c)) $\Omega=50, \omega=50, \lambda=1/4, \varepsilon=100, g=0.5, \nu=2$, one observes that a particle initially occupying the right well with spin-up undergoes spin-flip tunneling after time evolution; after a sufficiently long time the final total probability approaches a stable value of 1.15. However, the second case corresponds to three relations:
\[
\begin{aligned}
\text{(1)}\quad &J_3=J_4=0, &\quad\beta_r\beta_l&=J_1J_2;\\
\text{(2)}\quad &J_1=J_2=0, &\quad\beta_r\beta_l&=J_3J_4;\\
\text{(3)}\quad &J_1=J_2=J_3=J_4=0, &\quad\beta_l&=0.
\end{aligned}
\]
Taking (3) as an example, we plot in Fig. 7(d) the evolution probability with time. Here the quasi-energies are $E_{1}=E_{4}=0$ and $E_2=E_3=-\mathrm{i}\beta_r$. In this case the particle stays in its initial well; if the initial state is restricted to the other well, the final state still remains at its original position. Therefore, in this case the stability of the system depends on its initial state. For example, taking $\Omega=50, \omega=50, \lambda=1, \varepsilon=60.125, \nu=2$, for the initial condition $\ket{0,\uparrow}$ the total probability decays exponentially to zero with time, whereas for the initial condition $\ket{\downarrow ,0}$ the particle keeps its initial state, exhibiting coherent tunneling.

\begin{figure}[!htbp]
\centering
\includegraphics[width=0.65\linewidth]{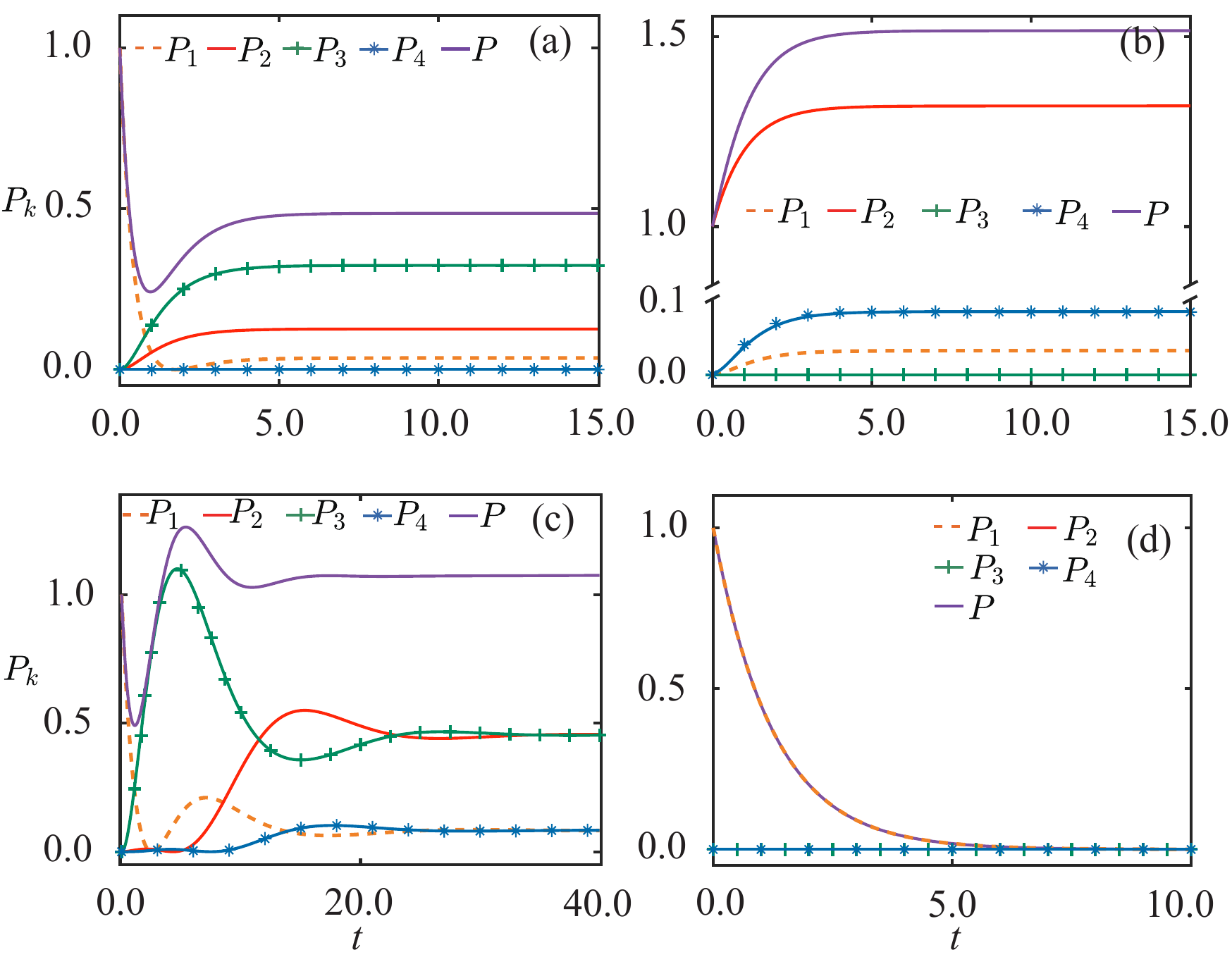}
\caption{In the unbalanced gain-loss nonreciprocal system, within the stable region, time evolution of each state probability $P_k=\lvert a_k(t)\rvert^2$ and of the total probability $P$ under different conditions. (a),(b) For even $\Omega/\omega$, the state evolution when the particle starts in the right and the left well, respectively; $\Omega=100, \lambda=0.3, \varepsilon=37.5$. (c), (d) For odd $\Omega/\omega$, the particle-probability evolution under different stable conditions; the particle starts in the right well with spin-up. The remaining parameters are $\nu=2, \omega=50, g=0.5$.}
\label{fig:7}
\end{figure}
\section{Conclusion}
In summary, this paper has studied the nonreciprocal Floquet tunneling dynamics of spin-orbit-coupled bosonic atoms in a non-Hermitian double-well potential. Based on conventional Floquet theory and the high-frequency approximation, we have analytically calculated the quasi-energies, the Floquet states, and the stability conditions of the non-Hermitian system. The results show that, in a nonreciprocal system with balanced gain and loss, continuous stable parameter regions appear when $\Omega/\omega$ is even. As the gain-loss strength increases, stable spin-flip tunneling is preferentially suppressed and the stable parameter region narrows. However, when $\Omega/\omega$ is odd, the previously continuous stable regions all become discrete. Thus, the size of the stable region depends on the nonreciprocal and balanced gain-loss strengths; whether the stable region is continuous or discrete depends on $\Omega/\omega$; and the particle's spin-flip tunneling can be tuned through the parameters $\lambda$ and $2\varepsilon/\omega$. On the other hand, when unbalanced gain and loss is introduced into the nonreciprocal system, we have also derived the balanced conditions for stable spin-flip tunneling when $\Omega/\omega$ is even and when it is odd. This study provides a new theoretical approach to realizing robust spin-flip tunneling and has important guiding significance for the stable transport of quantum states in non-Hermitian ultracold atomic systems.

\end{document}